\documentclass[preprint]{revtex4}
\usepackage{psfig}
\usepackage{graphicx}%
\usepackage{dcolumn}
\usepackage{amsmath}

\makeatletter
\def\btt#1{\texttt{\@backslashchar#1}}%
\DeclareRobustCommand\bblash{\btt{\@backslashchar}}%
\makeatother

\begin{document}

\title{Features of Motion Around  Global Monopole in Asymptotically dS/AdS Spacetime}
\
\author{Jian-gang Hao}

\author{Xin-zhou Li}\email{kychz@shtu.edu.cn}

\affiliation{SUCA, Shanghai United Center for Astrophysics,
Shanghai Normal University, Shanghai 200234, China
}%

\date{\today}

\begin{abstract}
Abstract : In this paper, we study the motion of test particle and
light around the Global Monopole in asymptotically dS/AdS
spacetime. The motion of a test particle and light in the exterior
region of the global monopole in dS/AdS spacetime has been
investigated. Although the test particle's motion is quite
different from the case in asymptotically flat spacetime, the
behaviors of light(null geodesic) remain unchanged except a
energy(frequency) shift. Through a phase-plane analysis, we prove
analytically that the existence of a periodic solution to the
equation of motion for a test particle will not be altered by the
presence of cosmological constant and the deficit angle, whose
presence only affects the position and type of the critical point
on the phase plane. We also show that the apparent capture section
of the global monopole in dS/AdS spacetime is quite different from
that in flat spacetime.
\end{abstract}

\pacs{04.40.-b, 11.10.Lm}

\maketitle

\textbf{PACS numbers}: 04.40.-b, 11.10.Lm

\vspace{0.4cm} \noindent\textbf{1. Introduction} \vspace{0.4cm}

Various of kinds of topological defects associated with the
spontaneous symmetry breaking (SSB) are very interesting objects,
whose only plausible production site is a cosmological phase
transition wherein they are produced by the Kibble mechanism.
\cite{vilenkin}. Domain walls are two-dimensional topological
defects, and strings are one-dimensional defects. Point-like
defects also arise in some theories which undergo SSB, and they
appear as monopoles. Global monopole, which has divergent mass in
flat spacetime, is one of the most important above mentioned
defects. The property of the global monopole in curved spacetime,
or equivalently, its gravitational effects, was firstly studied by
Barriola and Vilenkin\cite{barriola}. When one considers the
gravity, the linearly divergent mass of global monopole has an
effect analogous to that of a deficit solid angle plus that of a
tiny mass at the origin. Harari and Loust\`{o}\cite{harari}, and
Shi and Li\cite{li1} have shown that this small gravitational
potential is actually repulsive. A new class of cold stars,
addressed as D-stars(defect stars) have been proposed by Li
et.al.\cite{li2, li3}. One of the most important features of such
stars, comparing to Q-stars, is that the theory has monopole
solutions when the matter field is abscent, which makes the
D-stars behave very differently from the Q-stars. On the other
hand, there has been a renewed interest in AdS spacetime due to
the theoretical speculation of AdS/CFT correspondence, which state
that string theory in anti-de Sitter space (usually with extra
internal dimensions) is equivalent to the conformal field theory
in one less dimension\cite{maldacina, witten}. Recently, the
holographic duality between quantum gravity on de Sitter(dS)
spacetime and a quantum field theory living on the past boundary
of dS spacetime was proposed\cite{strominger} and the vortices in
dS spacetime was studied by Ghezelbash and Mann\cite{mann}. Many
authors conjectured that the dS/CFT correspondence bear a lot of
similarities with the AdS/CFT correspondence, although some
interpretive issues remain. The monopole and dyon solution in
gauge theories based on the various gauge group have been
found\cite{hooft}. However, in flat space there can not be static
soliton solution in the pure Yang-Mills theory\cite{deser}. The
presence of gravity can supply attractive force which binds
non-Abelian gauge field into a soliton. The cosmological constant
influence the behavior of the soliton solution significantly. In
asymptotically Minkowski spacetime the electric components are
forbidden in static solution\cite{volkov}. If the spacetime
includes the cosmological constant, forbidding the electric
components of the non-Abelian gauge fields fail, thus allowing
dyon solutions. A continuum of new dyon solutions in the
Einstein-Yang-Mills theory in asymptotically AdS spacetime have
been investigated\cite{bjoraker}, which are regular everywhere and
specified with their mass, and non-Abelian electric and magnetic
charges. Similarly, the presence of cosmological constant affects
the behavior of the global monopole remarkably. If the spacetime
is modified to include the positive cosmological constant, the
gravitation field of global monopole can be attractive in contrast
to the same problem in asymptotically Minkowski or AdS spacetime.

In a previous paper by Li and Hao\cite{hao}, it is shown that the
property of the global monopole in asymptotically dS/AdS spacetime
is very different from that in the asymptotically flat spacetime.
That is, the mass of the monopole might be positive in dS
spacetime if the cosmological constant is greater than a critical
value. In this paper, we will discuss the intriguing astrophysical
effects of global monopoles in some detail. That is, the behavior
of test particle and null geodesics in the exterior spacetime of
the global monopole in asymptotically dS/AdS sapcetime. With the
aid of phase-plane analysis, we carefully analyze the motion of
light and a test particle around a global monopole in dS/AdS
spacetime and show that the behavior is quite different from the
case for ordinary object in asymptotically flat spacetime.

\vspace{0.4cm} \noindent\textbf{2. The Global Monopole in
Asymptotically dS/AdS Spacetime}
 \vspace{0.4cm}

In this section, we briefly introduce the global monopole in
dS/AdS spacetime and one may refer to Ref.\cite{hao} for a more
completed description. The Lagaragian for the global monopole is
\begin{equation}\label{lag}
L=\frac{1}{2}\partial_\mu\phi^a\partial^\nu\phi^a-\frac{1}{4}\lambda^2(\phi^a\phi^a-\sigma_0)^2
\end{equation}

\noindent where $\phi^a$ is the triplet of Goldstone field and
possesses a internal O(3) symmetry. When the symmetry breaks down
to U(1), there will exist topological defects known as monopole.
The configuration describing monopole solution is

\begin{equation}\label{config}
  \phi^a=\sigma_0f(\rho)\frac{x^a}{\rho}
\end{equation}

\noindent where $x^ax^a=\rho^2$ and $a=1,2,3.$

When $f$ approaches unity at infinity, we will have a monopole
solution. The static spherically symmetric metric is

\begin{equation}\label{metric}
  ds^2=B(\rho)dt^2-A(\rho)d \rho^2-\rho^2(d\theta^2+ \sin^2\theta d\varphi^2)
\end{equation}

In dS/AdS spacetime, the equation of motion of the monopole field
and the Einstein equation are
\begin{equation}\label{scalareq}
  \frac{1}{A}f^{''}+[\frac{2}{Ar}+\frac{1}{2B}(\frac{B}{A})^{'}]f^{'}-\frac{2}{r^2}f
  -\lambda^2(f^2-1)f=0
\end{equation}

\begin{equation}\label{einsteineq}
G_{\mu\nu}+\beta g_{\mu\nu}=\kappa T_{\mu\nu}
\end{equation}

\noindent where $\beta$ is the cosmological constant and $\kappa=8
\pi G$. dS and AdS spacetime corresponds to the cases that $\beta$
is positive or negative respectively. By introducing the
dimensionless paramters $r=\sigma_0\rho$ and
$\epsilon^2=\kappa\sigma_0^2$, the Einstein equation can be
formally integrated and solution are as following

\begin{equation}\label{solution1}
 A^{-1}(r)=1-\epsilon^2+\frac{\beta}{3\sigma_0^2}r^2-\frac{2G\sigma_0M_A(r)}{r}
\end{equation}

\begin{equation}\label{solution2}
 B(r)=1-\epsilon^2+\frac{\beta}{3\sigma_0^2}r^2-\frac{2G\sigma_0M_B(r)}{r}
\end{equation}

\noindent where

\begin{eqnarray}\label{ma}
M_A(r)=4\pi\sigma_0\exp[-\triangle(r)]\times\int_0^r
dy\exp[\triangle(y)]
\{f^2-1+y^2[\frac{\lambda^2}{4}(f^2-1)^2+(1-\epsilon^2+\frac{\beta}{3\sigma_0^2}y^2)f^{'2}]\}
\end{eqnarray}

\noindent and

\begin{eqnarray}\label{mb}
M_B(r)=M_A(r)\exp[\widetilde{\triangle}(r)]
+\frac{r(1-\epsilon^2+\frac{\beta}{3\sigma_0^2}r^2)}{2}\{1-\exp[\widetilde{\triangle}(r)]\}
\end{eqnarray}

\noindent In which
\begin{equation}\label{delta}
\triangle(r)=\frac{\epsilon^2}{2}\int_0^rdy(yf^{'2})
\end{equation}

\noindent and

\begin{equation}\label{delta}
\widetilde{\triangle}(r)=\epsilon^2\int_\infty^rdy(yf^{'2})
\end{equation}

Next, we discuss the behavior of these functions in asymptotically
dS/AdS spacetime. A global monopole solution $f$ should approaches
unity when $r\gg1$. If this convergence is fast enough then
$M_A(r)$ and $M_B(r)$ will also quickly converge to finite values.
Therefore, we can find the asymptotic expansions:
\begin{equation}\label{asymf}
  f(r)=1-\frac{3\sigma_0^2}{\beta+3\lambda^2\sigma_0^2}\frac{1}{r^2}-
  \frac{9[2\beta\epsilon^2\sigma_0^4+3(2\epsilon^2-3)\lambda^2\sigma_0^6]}
  {2(2\beta-3\lambda^2\sigma_0^2)(\beta+3\lambda^2\sigma_0^2)^2}\frac{1}{r^4}+O(\frac{1}{r^6})
\end{equation}

\begin{equation}\label{asyma}
  M_A(r)=M_A+\frac{4\pi\sigma_0}{r}+O(\frac{1}{r^3})
\end{equation}

\begin{equation}\label{asymb}
  M_B(r)=M_A(r)(1-\frac{\epsilon^2}{r^4})+\frac{4\pi\sigma_0(1-\epsilon^2)}{r^3}+O(\frac{1}{r^7})
\end{equation}

\noindent where $M_A(\beta, \epsilon^2)\equiv \lim _{r\rightarrow
\infty}M_A(r)$, which is a function dependent on $\beta$ and
$\epsilon^2$.

We now investigate the motion of light and test particle around a
global monopole in asymptotically dS/AdS spacetime. Since the
effective mass $M_A(r)$ approaches very quickly its asymptotic
value, it is a good approximation to take it as the constant $M$,
unless we were interested in a test particle moving right into the
core of the monopole. THerefore, we consider the deodesic
equations in the metric (\ref{metric}) with

\begin{equation}\label{solution22}
 A(r)^{-1}=B(r)=1-\epsilon^2+\beta r^2-\frac{2G M}{r}
\end{equation}

\noindent where, for convenience, we have rescaled $\beta$ and $M$
as $\beta=\frac{\beta}{3\sigma_0}$ and $M=M\sigma_0$ respectively.

\vspace{0.4cm} \noindent\textbf{3. The Behavior of Null Geodesic
outside the Global Monopole in dS/AdS spacetime} \vspace{0.4cm}

The orbit of light outside of the monopole core can be obtained by
solving the geodesic equation
\begin{equation}\label{geodesic}
  \frac{d^2x^{\rho}}{d\tau^2}+\Gamma^{\rho}_{\alpha\beta}\frac{dx^{\alpha}}
  {d\tau}\frac{dx^{\beta}}{d\tau}=0
\end{equation}

\noindent where $\tau$ is the affine parameter. But by using the
fact that $g_{ab}\frac{dx^a}{d\tau}\frac{dx^b}{d\tau}=0$ for null
geodesics and the constant of motion
\begin{equation}\label{constant1}
 E=B(r)\frac{dt}{d\tau}
\end{equation}
\noindent and
\begin{equation}\label{constant2}
  L=r^2\frac{d\varphi}{d\tau}
\end{equation}

\noindent one can, considering the motion is confined on the
$\theta=\frac{\pi}{2}$ plane, easily obtain the equation for
geodesics as\cite{wald}:
\begin{equation}\label{nullgeodesic}
\frac{1}{2}\dot{r}^2+\frac{1}{2}B(r)(\frac{L^2}{r^2})=\frac{1}{2}E^2
\end{equation}

Introducing the effective potential
\begin{equation}\label{effv}
 V_{eff}=\frac{1}{2}B(r)(\frac{L^2}{r^2})=\frac{L^2\beta}{2}+
 \frac{L^2(1-\epsilon^2)}{2r^2}-\frac{L^2GM}{r^3}
\end{equation}

\noindent the geodesics of light becomes the same as a test
particle with unit mass moving in the effective potential Eq.(
\ref{effv}). It is not difficult to found that the effective
potential has a maximum
\begin{equation}\label{vmax}
 V_{max}=\frac{L^2\beta}{2}+\frac{L^2GM(1-\epsilon^2)^3}{2(3GM)^3}
\end{equation}

\noindent at
\begin{equation}\label{rcric}
r=r_c=\frac{3GM}{1-\epsilon^2}
\end{equation}

\noindent For convenience, we will choose $G=1$ in the following
discussion. When
\begin{equation}\label{bequ}
  \frac{1}{2}E^2=V_{max}
\end{equation}
\noindent we will have
\begin{equation}\label{bcrit}
  b_{crit}=\frac{3^{3/2}M}{\sqrt{27M^2\beta+(1-\epsilon^2)^3}}
\end{equation}

\noindent where $b_{crit}$ is the critical value of $b$, which is
the generalization of the apparent impact parameter and is defined
as $b=\frac{L}{E}$. In the case that the mass of the monopole is
positive, that is, the spacetime is asymptotically de Sitter and
the cosmological constant is greater than a critical value, the
capture cross section of the monopole will be
\begin{equation}\label{crosssection}
\sigma_c=\pi b_{crit}^2=\frac{27\pi
M^2}{27M^2\beta+(1-\epsilon^2)^3}
\end{equation}

It is not difficult to prove that the path of light has a turning
point at the largest radius,$R_0$, where
$\frac{dr}{d\varphi}|_{r=R_0}=0$. Using
Eqs.(\ref{constant1})-(\ref{nullgeodesic}), one can obtain the
relation between $R_0$ and $b$ as
\begin{equation}\label{R0}
R_0=2b\sqrt{\frac{1-\epsilon^2}{3(1-b^2\beta)}}\cos\{\frac{1}{3}
\arccos[-(\frac{3}{1-\epsilon^2})^{3/2}\frac{M}{b}\sqrt{1-b^2\beta}]\}
\end{equation}

\noindent Obviously, $R_0=b$ when $M=\epsilon=\beta=0$. In order
to compute the deflect angle of the light, we rewrite the
Eqs.(\ref{constant1})-(\ref{nullgeodesic}) as
\begin{equation}\label{deflight}
 \frac{d\varphi}{dr}=[r^4b^{-2}-r(r-\epsilon^2 r+\beta
 r^3-2M)]^{-1/2}
\end{equation}
\noindent The change of light when passing a monopole should be
$\Delta\varphi=\varphi_\infty-\varphi_{-\infty}$. Considering the
symmetry, we have
\begin{equation}\label{angle}
\Delta\varphi=2\int_{R_0}^\infty\frac{dr}{[r^4b^{-2}-r(r-\epsilon^2
r+\beta r^3-2M)]^{1/2}}
\end{equation}

\begin{figure}
\psfig{file=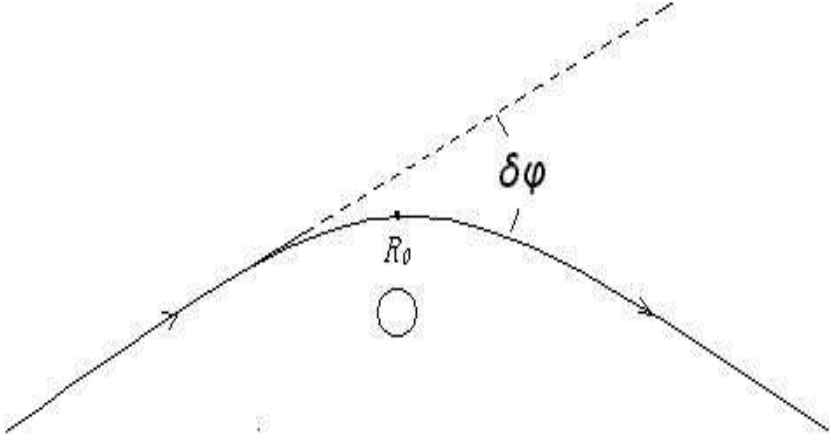,height=4in,width=5in} \caption{Diagram
illustrating the "bending of light" effect}
\end{figure}

\noindent The deflection of light up to the first order of $M$
($M$ is small in the unit $G=1$) is given by (See Fig.1 for the
definition of $ \delta\varphi$):
\begin{equation}\label{dangle}
 \delta\varphi=\Delta\varphi-\pi\approx M\frac{\partial(\Delta\varphi)}{\partial
 M}|_{M=0}=\frac{4M}{(1-\epsilon^2)^{3/2}R_0}
\end{equation}

\noindent In terms of the apparent impact parameter $b$, the
Eq.(\ref{dangle})
\begin{equation}\label{dangleb}
 \delta\varphi=\frac{2M\sqrt{3(1-b^2\beta)}}{b(1-\epsilon^2)^2\cos\{\frac{1}{3}
\arccos[-(\frac{3}{1-\epsilon^2})^{3/2}\frac{M}{b}\sqrt{1-b^2\beta}]\}}
\end{equation}

\noindent One can easily find that when setting
$\beta=\epsilon=0$, the above Eqs(\ref{dangle}), (\ref{dangleb})
will reduce to the well known form for schwarzschild spacetime.
From Eq.(\ref{dangleb}), we can find that for a beam of light with
a specific $b$ or $L/E$, its deflecting behavior will be
significantly influenced by the presence of the deficit angle
$\epsilon^2$ and the cosmological constant $\beta$, which could be
employed as an characteristic feature of the monopole in dS/AdS
spacetime.

On the other hand, from Eq.(\ref{nullgeodesic}), (\ref{effv}), if
we redefine $E$ as $\widetilde{E}^2=E^2-L^2\beta$, the effects of
the cosmological constant on the motion of light will be ascribed
to a shift of its energy. But, as we will show later, this
property does not hold true for the timelike test particles.

\vspace{0.4cm} \noindent\textbf{4. The Motion of Timelike
Particles Around a Global Monopole in dS/AdS spacetime}
 \vspace{0.4cm}

From Eqs(\ref{geodesic})-(\ref{constant2}), the equation of motion
for a timelike test particle can be expressed as
\begin{equation}\label{timelike}
 (\frac{L}{r^2}\frac{dr}{d\varphi})^2=E^2-\mu^2B(r)-\frac{L^2}{r^2}B(r)
\end{equation}
\noindent where $\mu=\frac{\mu}{\sigma_0}$ and
$E=\frac{E}{\sigma_0}$ are the rescaled mass and energy of the
test particle. Introducing $\chi=\frac{1}{r}$, substituting it
into Eq.(\ref{timelike}) and then differentiating the equation
with respect to $\varphi$, one will obtain the following equation
\begin{equation}\label{timelikenew}
\frac{d^2
\chi}{d\varphi^2}=\frac{\gamma}{\chi^3}+\frac{1}{p}-(1-\epsilon^2)\chi+\alpha\chi^2
\end{equation}

\noindent where $\alpha$, $p$ and $\gamma$ are three dimensionless
parameters defined as:

\begin{eqnarray}
\alpha &=& 3GM\\\nonumber
\frac{1}{p}&=&\frac{GM\mu^2}{L^2}\\\gamma&=&\frac{\mu^2\beta}{L^2}\nonumber
\end{eqnarray}

Noting that Eq.(\ref{timelikenew}) is a nonlinear differential
equation, it can be integrated formally as
\begin{equation}\label{inte}
\varphi-\varphi_0=\int_{\chi_0}^{\chi}\frac{d\chi}{\sqrt{\int_{\chi_0}^{\chi}2
[\frac{\gamma}{\chi^3}+\frac{1}{p}-(1-\epsilon^2)\chi+\alpha\chi^2]d\chi+\dot{\chi}_0^2}}
\end{equation}
\noindent where, $\dot{\chi}_0^2$ is the initial value of
$\frac{d\chi}{d\varphi}|_{\varphi=\varphi_0}$. However, it is
impossible to obtain the exact expression by integrating the above
equation. In the following, we will gain some qualitative property
of the system with the aid of phase-plane analysis without solving
the equation numerically. To do so, we introduce two parameters as
$x=\chi$ and $y=\frac{d\chi}{d\varphi}$ and the autonomous system
corresponding to Eq.(\ref{timelikenew}) will be

\begin{eqnarray}\label{autos}
\frac{dx}{d\varphi} &=& f(x,y)=y\\\nonumber \frac{dy}{d\varphi}&=&
g(x,y)=\frac{\gamma}{x^3}+\frac{1}{p}-(1-\epsilon^2)x+\alpha x^2
\end{eqnarray}

Now, we prove the existence of periodic solution. According to the
well known Bendixson's criterion\cite{bendixson}, the equation of
motion will have periodic solution if the divergence of the
functional vector of the autonomous system is vanishing, i.e.,
$\nabla\cdot(f, g)=0$. It is obvious that the functional vector
$(f, g)$ corresponding to the Eqs.(\ref{autos}) satisfies the
criterion and therefore indicates that Eq.(\ref{timelikenew}) has
a periodic solution. One can also found that the periodic solution
exists when $\gamma=\epsilon=0$ which is the case of an ordinary
star in asymptotically flat spacetime. This shows that the
presence of the cosmological constant and deficit angle will not
exclude the existence of periodic solution from the equation of
motion.

Next, we analyze the critical points on the phase plane. The
critical point is $(x_0, 0)$, where $x_0$ satisfies

\begin{equation}\label{critialp}
\frac{\gamma}{x_0^3}+\frac{1}{p}-(1-\epsilon^2)x_0+\alpha x_0^2=0
\end{equation}

\noindent To analyze the type of the critical point, we firstly
linearize the Eq.(\ref{autos}) and then do the translation
$x=x-x_0$. Thus the linearized equations should be:

\begin{eqnarray}\label{lautos}
\frac{dx}{d\varphi} &=& y\\\nonumber \frac{dy}{d\varphi}&=& \delta
x
\end{eqnarray}

\noindent where
\begin{equation}\label{delta}
\delta=-\frac{3\gamma}{x_0^4}-(1-\epsilon^2)+2\alpha x_0
\end{equation}

\noindent Using Eqs.(\ref{critialp}) and (\ref{delta}) could be
rewritten as

\begin{equation}\label{delta1}
\delta=\frac{3}{px_0}-4(1-\epsilon^2)+5\alpha x_0
\end{equation}

The eigenvalues corresponding to the system of equations will be
\begin{equation}\label{eigenv}
\lambda_{1, 2}=\pm\sqrt{\delta}
\end{equation}

\noindent The types of the critical point could be classified
according to the eigenvalues as following:

\noindent \textbf{I}. when $\delta>0$, we have
$\lambda_1<0<\lambda_2$, which indicates that the critical point
is an unstable saddle point. Considering the Eq.(\ref{delta1}),
this case will correspond to the condition that

\noindent (1). $\alpha>0$ and $\Delta<0$.

\noindent (2). $\alpha>0$, $\Delta>0$ and
$x_0>\frac{2(1-\epsilon^2)+\sqrt{\Delta}}{5\alpha}$ or
$0<x_0<\frac{2(1-\epsilon^2)-\sqrt{\Delta}}{5\alpha}$.

\noindent (3). $\alpha<0$, $\Delta>0$ and
$\frac{2(1-\epsilon^2)-\sqrt{\Delta}}{5\alpha}<x_0<\frac{2(1-\epsilon^2)+\sqrt{\Delta}}{5\alpha}$,
where $\Delta=(1-\epsilon^2)^2-\frac{15\alpha}{4p}$.

\noindent \textbf{II}. when $\delta<0$, we have two pure imaginary
eigenvalues $\lambda_{1, 2}=\pm\textbf{i}\sqrt{|\delta|}$, which
indicates that the critical point is stable center. Considering
the Eq.(\ref{delta1}), this case will correspond to the condition
that

\noindent (1). $\alpha<0$ and $\Delta<0$.

\noindent (2). $\alpha>0$, $\Delta>0$ and
$\frac{2(1-\epsilon^2)-\sqrt{\Delta}}{5\alpha}<x_0<\frac{2(1-\epsilon^2)+
\sqrt{\Delta}}{5\alpha}$.

\noindent (3). $\alpha<0$, $\Delta>0$ and
$x_0>\frac{2(1-\epsilon^2)+\sqrt{\Delta}}{5\alpha}$ or
$0<x_0<\frac{2(1-\epsilon^2)-\sqrt{\Delta}}{5\alpha}$.

\noindent What we need to point out is that the case $\alpha<0$ is
possible because the mass of the global monopole could be
negative.

\noindent \textbf{III}. when $\delta=0$, we have $\lambda_{1,
2}=0$, which, together with the form of the autonomous system
Eqs.(\ref{lautos}), indicates that the motion is uniformly on the
lines $y=Constants$ and all the the points on the lines
$y=Constants$ are balanced positions. In Fig.2, Fig.3 and Fig.4,
we show the phase graph for different initial values and different
parameters.

\begin{figure}
\psfig{file=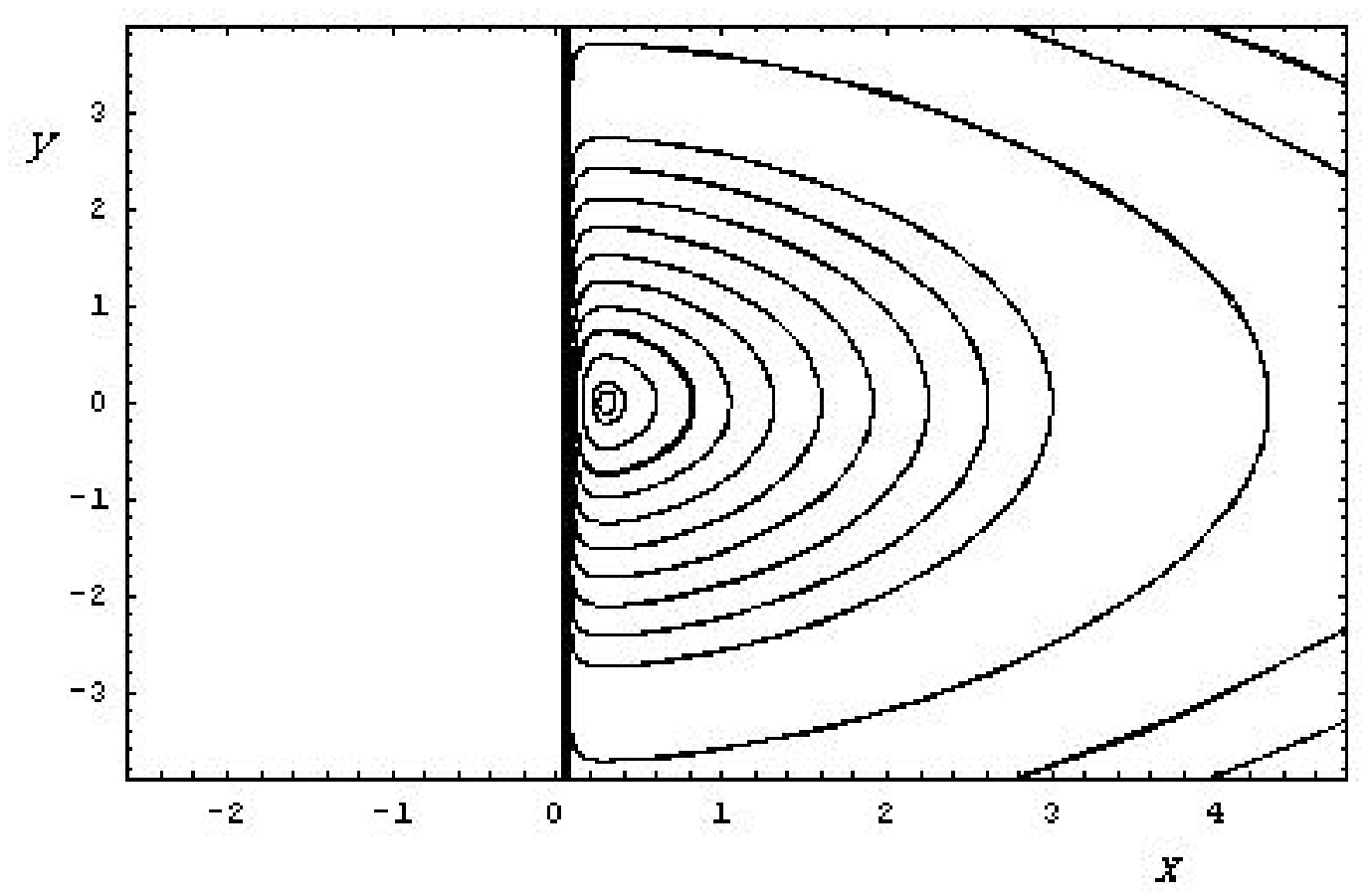,height=3in,width=4in}
\caption{ the phase
graph when $\alpha=0.10$, $\gamma=0.01$, $\epsilon=0.01$ and
$p=11$}
\psfig{file=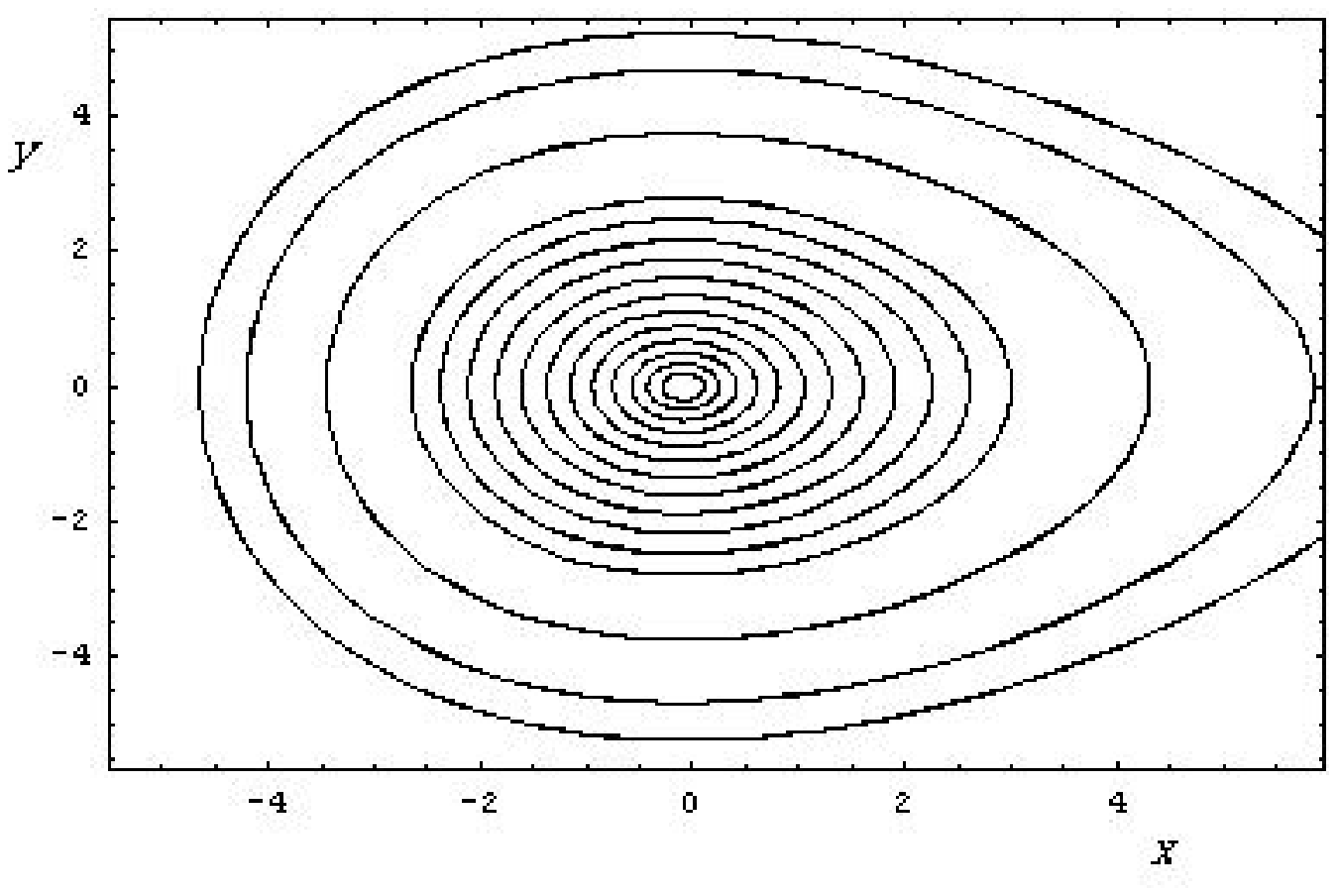,height=3in,width=4in}
\caption{ the
phase graph when $\alpha=0.10$, $\gamma=0.00$, $\epsilon=0.01$ and
$p=11$}
\psfig{file=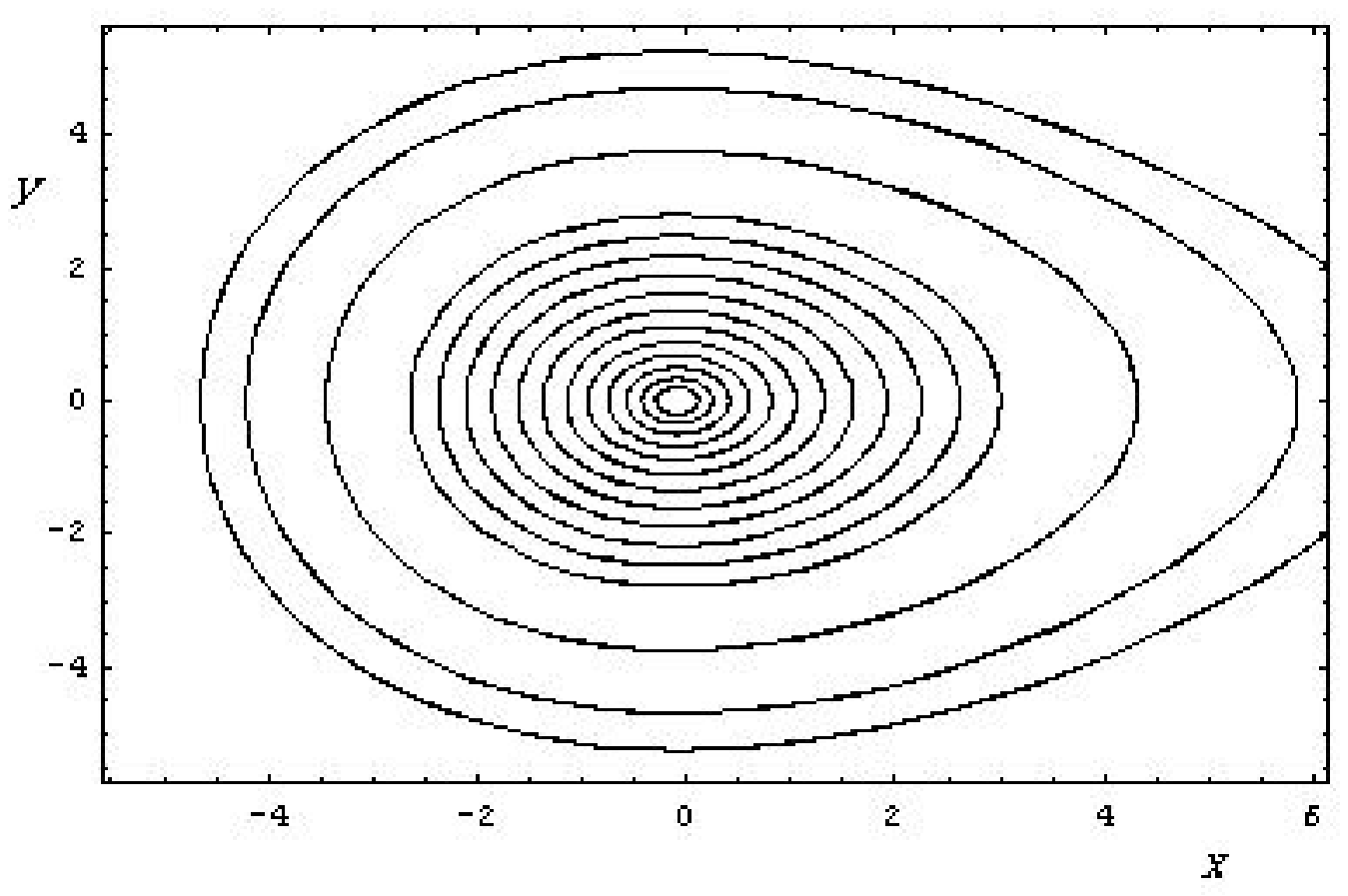,height=3in,width=4in}
\caption{ the
phase graph when $\alpha=0.10$, $\gamma=0.00$, $\epsilon=0.00$ and
$p=11$}
\end{figure}

Hitherto, we consider only the qualitative property of the
equation of motion of test particle in the exterior region of a
global monopole in dS/AdS spacetime. One can simply perform some
numerical calculation under different parameters so as to get the
solutions. But this is beyond the scope of this paper and is not
the very purpose of this paper. In the following, we will study
the precession of the test particle around the global monopole.
Noting that the cosmological constant is generally very small, the
term $\frac{\gamma}{\chi^3}$ could be neglected when $\chi=r^{-1}$
is very large, which agrees with our notion that the presence of
cosmological will not change the property of Einsteinian gravity
at the scale of solar system. Therefore, when we consider the
behavior of a test particle around a global monopole, we can
neglect the influence from the cosmological constant term. So, the
equation of motion for a test particle now reduces to
\begin{equation}\label{timelikenew1}
\frac{d^2
\chi}{d\varphi^2}=\frac{1}{p}-(1-\epsilon^2)\chi+\alpha\chi^2
\end{equation}

It is obvious that when one neglects the higher order term
$\chi^2$ and sets $\epsilon=0$, the resulting orbit of the test
particle obtained from the above equation will be the case
attained through Newtonian Mechanics
\begin{equation}\label{newtons}
\chi=\frac{1}{p}(1+e\cos\varphi)
\end{equation}
\noindent where $e$ is the eccentricity. In order to get the
solution of Eq.(\ref{timelikenew1}), as done in
Ref.(\cite{landau}), we decompose $\chi$ as $\chi=\chi_0+\chi_1$.
\noindent Then we have
\begin{equation}\label{xi0}
 \chi_0=\frac{1}{(1-\epsilon^2)p}[1+e\cos(\sqrt{1-\epsilon^2}\varphi)]
\end{equation}

\begin{equation}\label{xi1}
 \chi_1=\frac{\alpha e}{(1-\epsilon^2)p^2}(\sqrt{1-\epsilon^2}\varphi)\sin(\sqrt{1-\epsilon^2}\varphi)
\end{equation}

Obviously, when $\frac{\alpha}{p}\varphi\ll1$, we can write the
solution as:
\begin{equation}\label{solutionx}
\chi=\chi_0+\chi_1 \simeq \frac{1}{p}\{1+e
\cos[\sqrt{1-\epsilon^2}(1-\frac{\alpha}{(1-\epsilon^2)p})\varphi]\}
\end{equation}

Now, it is straightforward to estimate the precession of a test
particle when it rotate one loop around the global monopole. It
will take the test particle $\frac{2\pi}{\sqrt{1-\epsilon^2}}$ to
complete one loop. The precession after one loop will be
\begin{equation}\label{precession}
\delta\varphi=\frac{6\pi GM}{a(1-\epsilon^2)(1-e^2)}
\end{equation}
\noindent where $a=\frac{p}{1-e^2}$. Comparing this result with
the familiar result for the precession around sun, one can find
that the there would be an modification. This modification can be
more clear if we rewrite Eq.(\ref{precession}) as
\begin{equation}\label{precessionnew}
\delta\varphi=\frac{6\pi GM}{a(1-e^2)}+\frac{6\pi GM
\epsilon^2}{a(1-e^2)}
\end{equation}
\noindent This show that the test particle around the global
monopole will have an extra precession $\frac{6\pi GM
\epsilon^2}{a(1-e^2)}$ than that around a ordinary star.

\vspace{0.4cm} \noindent\textbf{4. Discussion}
 \vspace{0.4cm}

In this paper, we study the motion of test particle and light
around the global monopole in dS/AdS spacetime. We show that the
null geodesics and timelike geodesics behave very differently in
the presence of cosmological constant and deficit angle. The
motion of light is not drastically changed except a factor of
$(1-\epsilon^2)$ and a shift of its energy, which consequently
lead to different relation between the apparent impact parameter
and the deflect angle.

However, the behavior of timelike geodesics has been affected
significantly when there are cosmological constant and deficit
angle. By using the phase-plane analysis, we investigate the
qualitative property of the dynamical equation governing the
motion of a test particle around the global monopole in
asymptotically dS/AdS spacetime. We prove that the equation of
motion possesses periodic solution. This property is not altered
by the cosmological constant and deficit angle, which, however,
affect the position of the critical point and its type on the
phase plane. The conditions under which the critical point is
stable center and unstable saddle point respectively have been
given too. The precession of the test particle around a global
monopole has been investigated and the influence of deficit angle
on the precession has been manifested.

It is very interesting to consider a further generalization of
this model by requiring that the global monopole possess a U(1)
charge. This could be realized by coupling to the monopole field
with a scalar field with local U(1) symmetry. For details, see
Ref.(\cite{li4,li3}). In this case, the effective
potential(\ref{effv})becomes
\begin{equation}\label{charev}
 V_{eff}(r)=
 \begin{cases}
  (\frac{1}{2}(1-\epsilon^2+\beta r^2-\frac{2G M}{r}+\frac{Q^2}{r^2})(\frac{L^2}{r^2}+1) & , \text{for time-like geodesics}\\
   (\frac{1}{2}(1-\epsilon^2+\beta r^2-\frac{2G
   M}{r}+\frac{Q^2}{r^2})\frac{L^2}{r^2}&, \text{for null
   geodesics}
 \end{cases}
\end{equation}
\noindent If we set $\epsilon=0$, this will reduce to the
Reissner-Nordstr\"{o}m metric in asymptotically dS/AdS spacetime,
which has been thoroughly studied in Ref.\cite{stuchlik}. The
Introduction of the deficit angle affects the asymptotically
dS/AdS Reissner-Nordstr\"{o}m spacetime in a similar fashion as it
does on the asymptotically dS/AdS Schwarzschild spacetime as we
discussed in the former part of this paper and we will not repeat
it again.

\vspace{0.8cm} \noindent ACKNOWLEDGMENTS

This work was partially supported by National Nature Science
Foundation of China under Grant No. 19875016, National Doctor
Foundation of China under Grant No. 1999025110, and Foundation of
Shanghai Development for Science and Technology under Grant No.
01JC14035.

\end{document}